\documentclass[twocolumn]{jpsj3}

\usepackage{txfonts}
\usepackage{bm}

\title{Spin-Triplet Pairing State of Sr$_2$RuO$_4$ in the $c$-Axis Magnetic Field}

\author{Shuhei TAKAMATSU$^{1}$\thanks{E-mail:~takamatsu@phys.sc.niigata-u.ac.jp} and
Youichi YANASE$^{1,2}$}
\inst{$^1$Graduate School of Science and Technology, Niigata University, Niigata, 950-2181, Japan\\
$^2$Department of Physics, Niigata University, Niigata, 950-2181, Japan} 

\abst{We investigate the spin-triplet superconducting state of Sr$_2$RuO$_4$ 
in the magnetic field along the {\it c}-axis on the basis of the four-component
Ginzburg-Landau~(GL) model with a weak spin-orbit coupling.
We consider superconducting states described by the d-vector parallel to the {\it ab}-plane 
(${\bm d}\parallel ab$), and find that three spin-triplet pairing states
are stabilized in the magnetic field-temperature~($H$-$T$) phase diagram. 
Although a helical state is stable at low magnetic fields, a chiral II state 
is stabilized at high magnetic fields. 
A non-unitary spin-triplet pairing state appears near the transition temperature owing to the 
coupling of magnetic field and chirality. 
We elucidate synergistic and/or competing roles of the magnetic field, chirality, and 
spin-orbit coupling. 
It is shown that a fractional vortex lattice is stabilized in the chiral II phase
owing to the spin-orbit coupling.} 

\kword{Spin-triplet superconductivity, spin-orbit coupling, fractional vortex, Sr$_2$RuO$_4$}

\begin{document}
\maketitle
Since the discovery of superconductivity in Sr$_2$RuO$_4$ by Maeno {\it et~al}.,\cite{Maeno_et_al}
extensive studies have been carried out to clarify the symmetry of superconductivity in Sr$_2$RuO$_4$. 
It has been shown that Sr$_2$RuO$_4$ is a spin-triplet superconductor from both theoretical
and experimental points of view.\cite{Mackenize_et_al,Maeno_et_al_jpsj}
For instance, the Knight shift measurement by nuclear magnetic resonance (NMR) experiments is
one of the convincing piece of evidence of the spin-triplet pairing state.
The NMR measurements showed the temperature-independent Knight shift through the superconducting transition 
in both field directions along the {\it ab}-plane\cite{Ishida_et_al} and along
the {\it c}-axis.\cite{Murakawa_et_al} 
These results indicate a spin-triplet superconducting state with
a d-vector perpendicular to the magnetic field.
This means that the ``spin-orbit coupling'' of spin-triplet Cooper pairs is very weak in Sr$_2$RuO$_4$. 
According to the experiments of muon spin rotation~($\mu$SR)\cite{G_M_Luke_musr} and 
the Kerr effect,\cite{Jing_kerr}~the time reversal symmetry is spontaneously broken in 
the superconducting state. 
These findings indicated that the chiral spin-triplet pairing state described by the d-vector
${\bm d}=(p_{\rm x}\pm{\rm i}p_{\rm y}){\hat z}$ is stabilized at zero magnetic field. 

When the spin-orbit coupling is weak as indicated by NMR measurement\cite{Ishida_et_al,Murakawa_et_al} 
and by theoretical estimation\cite{Yanase_Ogata}, 
the d-vector is rotated by a magnetic field along the {\it c}-axis 
so as to be parallel to the {\it ab}-plane. 
However, most theoretical studies focused on the chiral state
${\bm d}=(p_{\rm x}\pm{\rm i}p_{\rm y}){\hat z}$
\cite{Agterberg,Agterberg2,Kita_PRL83_1846,Kato,Kato-Hayashi,Takigawa,Ichioka,Garaud,Silaev,Huo}
assuming that a strong spin-orbit coupling pins the direction of the d-vector. 
It is highly desired to investigate pairing states in the {\it c}-axis magnetic field 
by taking a weak spin-orbit coupling into account, but no such theoretical study has yet been conducted.
The purpose of our study is to determine the spin-triplet superconducting state for $H \parallel c$ 
in the presence of a weak spin-orbit coupling. 

Several issues regarding the superconductivity in Sr$_2$RuO$_4$ remain unsettled. 
For instance, the chiral edge state\cite{Matsumoto-Sigrist} has not been observed\cite{Kallin2012}, 
and the first-order superconducting transition at high magnetic fields along the {\it ab}-plane~\cite{Yonezawa} 
is incompatible with the weak-coupling theory for spin-triplet superconductors. 
To resolve these unresolved issues, our verification of superconducting phases 
for $H \parallel c$ will provide a basis for discussion. 
Our study would also give an insight into the spin-triplet superconducting state of 
UPt$_3$\cite{Joynt}, in which a weak spin-orbit coupling has been indicated\cite{Tou}. 

We here construct a four-component GL model assuming a weak spin-orbit coupling. 
Although the chiral state ${\bm d}=(p_{\rm x}\pm{\rm i}p_{\rm y}){\hat z}$ may be stabilized 
at zero magnetic field\cite{Mackenize_et_al,Maeno_et_al_jpsj},  
we assume that the d-vector is parallel to the {\it ab}-plane in the magnetic field along the {\it c}-axis 
so as to gain the magnetic energy. 
Focusing on this high-field superconducting phase,
we take into account two spin components of order parameters,
namely, ${\bm d} \parallel \hat{x}$ and ${\bm d} \parallel \hat{y}$. 
The D$_{\rm 4h}$ symmetry of the crystal structure of Sr$_2$RuO$_4$ allows two orbital components 
of $p$-wave order parameters, namely, $p_{\rm x}$ and $p_{\rm y}$, which are equivalently 
denoted by the chirality, that is, $p_{\rm x} \pm {\rm i} p_{\rm y}$. 
These $2 \times 2 = 4$ component order parameters correspond to 
four irreducible representations in the D$_{\rm 4h}$ symmetry, 
${\bm d}=p_{\rm x}{\hat x}\pm p_{\rm y}{\hat y}$ and 
${\bm d}=p_{\rm y}{\hat x}\pm p_{\rm x}{\hat y}$\cite{Sigrist_Ueda}.  
These pairing states are nearly degenerate when we assume a weak spin-orbit coupling. 
For convenience, we describe the order parameters as 
$\Delta_{\sigma\sigma}({\bm r},{\bm k})=\Delta_{\sigma\sigma,{\rm x}}({\bm r})\phi_{\rm x}({\bm k})
+\Delta_{\sigma\sigma,{\rm y}}({\bm r})\phi_{\rm y}({\bm k})$ 
using four component order parameters, $(\Delta_{\uparrow\uparrow,{\rm x}},\Delta_{\uparrow\uparrow,{\rm y}}
,\Delta_{\downarrow\downarrow,{\rm x}},\Delta_{\downarrow\downarrow,{\rm y}})$. 
$\phi_{\rm x}({\bm k})$ and $\phi_{\rm y}({\bm k})$ stand for pairing functions with 
the same symmetry as $k_{\rm x}$ and $k_{\rm y}$ under the D$_{\rm 4h}$ point group, respectively. 

The GL free-energy density is given by
\begin{equation}
  f=\sum_{\sigma=\uparrow,\downarrow}(f^{0}_{\sigma}+f^{\rm SO1}_{\sigma})+f^{\rm SO2},
\end{equation}
where $f^{0}_{\sigma}$ is the same as the GL free-energy density for the E$_{\rm u}$ representation
of the tetragonal point group\cite{Sigrist_Ueda},
\begin{align}
  f^{0}_{\sigma}=&\alpha\bigl(|\Delta_{\sigma\sigma,{\rm x}}|^2+|\Delta_{\sigma\sigma,{\rm y}}|^2\bigr)
  +\frac{\beta_1}{2}\bigl(|\Delta_{\sigma\sigma,{\rm x}}|^2+|\Delta_{\sigma\sigma,{\rm y}}|^2\bigr)^2 \nonumber\\
  &+\frac{\beta_2}{2}(\Delta_{\sigma\sigma,{\rm x}}\Delta^*_{\sigma\sigma,{\rm y}}-{\rm c.c.})^2 
  +\beta_3|\Delta_{\sigma\sigma,{\rm x}}|^2|\Delta_{\sigma\sigma,{\rm y}}|^2 \nonumber\\
  &+\xi^2_1\bigl[|D_{\rm x}\Delta_{\sigma\sigma,{\rm x}}|^2
    +|D_{\rm y}\Delta_{\sigma\sigma,{\rm y}}|^2\bigr] \nonumber\\
  &+\xi^2_2\bigl[|D_{\rm x}\Delta_{\sigma\sigma,{\rm y}}|^2
    +|D_{\rm y}\Delta_{\sigma\sigma,{\rm x}}|^2\bigr] \nonumber\\
  &+\xi^2_3\Bigl\{\bigl[(D_{\rm x}\Delta_{\sigma\sigma,{\rm x}})(D_{\rm y}\Delta_{\sigma\sigma,{\rm y}})^*
    +{\rm c.c.}\bigr] \nonumber\\
  &+\bigl[(D_{\rm x}\Delta_{\sigma\sigma,{\rm y}})(D_{\rm y}\Delta_{\sigma\sigma,{\rm x}})^*
    +{\rm c.c.}\bigr]\Bigr\}.
  \label{eq:f0}
\end{align}
We adopt $\alpha=\alpha_0(T-T^0_{\rm c})$, where $T^0_{\rm c}$ is the transition temperature
at zero magnetic field in the absence of spin-orbit coupling.
We use the conventional notation
$D_{j}=\nabla_{j}+(2\pi{\rm i}/\varPhi_{0})A_{j}$ and $\varPhi_{0}=hc/2|e|$.
In the weak coupling theory, parameters satisfy the relations following  
$\beta_2/\beta_1 = \langle|\phi_{\rm x}|^2|\phi_{\rm y}|^2\rangle/\langle|\phi_{\rm x}|^4\rangle$,
$\xi^2_2/\xi^2_1=\langle|\phi_{\rm x}|^2 v^2_{\rm y}\rangle/\langle|\phi_{\rm x}|^2 v^2_{\rm x}\rangle$,
and $\xi^2_3/\xi^2_1=\langle\phi_{\rm x}\phi^*_{\rm y} v_{\rm x}v_{\rm y}\rangle/\langle|\phi_{\rm x}|^2 v^2_{\rm x}\rangle$, 
where $v_{\rm x}$ and $v_{\rm y}$ are the Fermi velocity and 
brackets $\langle\rangle$ denote the average over the Fermi surface. 
When we assume $\left(\phi_{\rm x}({\bm k}\right), \phi_{\rm y}({\bm k})) 
= \left(v_{\rm x}({\bm k}), v_{\rm y}({\bm k})\right)$, we obtain $\xi_1 > \xi_2 = \xi_3$. 
On the other hand, we find $\xi_1 < \xi_2$ for the pairing functions obtained by a perturbation theory 
for the Hubbard model\cite{Udagawa_Yanase_Ogata}.  

A weak spin-orbit coupling is taken into account in the quadratic terms 
$f^{\rm SO1}_{\sigma}$ and $f^{\rm SO2}$, which are given by
\begin{gather}
  f^{\rm SO1}_{\sigma}=
  \sigma\epsilon({\rm i}\Delta_{\sigma\sigma,{\rm x}}\Delta^*_{\sigma\sigma,{\rm y}}+{\rm c.c.}),
  \label{eq:fso1}
  \\
  f^{\rm SO2}=\delta\bigl[(\Delta_{\uparrow\uparrow,{\rm x}}\Delta^*_{\downarrow\downarrow,{\rm x}}
    -\Delta_{\uparrow\uparrow,{\rm y}}\Delta^*_{\downarrow\downarrow,{\rm y}})+{\rm c.c.}\bigr].
  \label{eq:fso2}
\end{gather}
According to a theoretical analysis based on the three-orbital Hubbard model 
for Sr$_2$RuO$_4$\cite{Yanase_Ogata},~the $f^{\rm SO1}_{\sigma}$ term arises from the $(\alpha,\beta)$-bands,
while the $f^{\rm SO2}$ term is due to the coupling between the $(\alpha,\beta)$-bands
and the $\gamma$-band.

We determine the pairing state by minimizing the above GL free-energy density 
using the variational method. 
We take into account variational wave functions of Cooper pairs so that the solution of
the linearized GL equation is reproduced.
Writing $\Delta_{\sigma\sigma,1}=(\Delta_{\sigma\sigma,{\rm x}}-{\rm i}\Delta_{\sigma\sigma,{\rm y}})/\sqrt{2}$
and
$\Delta_{\sigma\sigma,2}=(\Delta_{\sigma\sigma,{\rm x}}+{\rm i}\Delta_{\sigma\sigma,{\rm y}})/\sqrt{2}$, 
and differentiating the quadratic terms of eq.~(\ref{eq:f0}) with respect to 
$\Delta_{\sigma\sigma,1}$ and $\Delta_{\sigma\sigma,2}$, we obtain the linearized GL equation 
in the absence of spin-orbit coupling as 
\begin{align}
  \frac{2\pi H}{\varPhi_0}
  \left(
  \begin{array}{cc}
    \kappa(1+2\Pi_+\Pi_-) & -\rho_{-}\Pi^2_+-\rho_{+}\Pi^2_- \\
    -\rho_{+}\Pi^2_+-\rho_{-}\Pi^2_-  & \kappa(1+2\Pi_+\Pi_-)
  \end{array}
  \right)
  \left(
  \begin{array}{c}
    \Delta_{\sigma\sigma,1} \\
    \Delta_{\sigma\sigma,2} 
  \end{array}
  \right)  \nonumber \\
  =2\lambda
  \left(
  \begin{array}{c}
    \Delta_{\sigma\sigma,1} \\
    \Delta_{\sigma\sigma,2} 
  \end{array}
  \right),
  \label{eq:tenmoto}
\end{align}
where $\kappa=\xi^2_1+\xi^2_2$, $\rho_{\pm}=\xi^2_1-\xi^2_2\pm2\xi^2_3$, and 
$\Pi_{\pm}=-(\varPhi_0/4\pi H)^{1/2}({\rm i}D_{\rm y}\pm D_{\rm x})$.
Note that this equation is independent of the spin label $\sigma$. 

For $\xi_1>\xi_2$, 
the minimum eigenvalue $\lambda = \lambda_{\rm min}$ is obtained for the pairing state 
with a positive chirality, in which the order parameters
$(\Delta_{\sigma\sigma,1}, \Delta_{\sigma\sigma,2})=(\psi_{1+}({\bm r},{\bm \delta}), \psi_{2+}({\bm r},{\bm \delta}))$ 
are described using the Landau level expansion 
\begin{equation}
  \left(
  \begin{array}{c}
    \psi_{1+}({\bm r},{\bm \delta}) \\
    \psi_{2+}({\bm r},{\bm \delta})
  \end{array}
  \right)=
  \sum_{n\geq 0}\left(
  \begin{array}{c}
    a_{4n} \varphi_{4n}({\bm r},{\bm \delta}) \\
    a_{4n+2} \varphi_{4n+2}({\bm r},{\bm \delta})
  \end{array}
  \right). 
  \label{eq:smallest_eigenstate}
\end{equation}
The superconducting transition temperature in the absence of spin-orbit coupling is obtained as 
$T^{0}_{\rm  c}(H) = T^{0}_{\rm c}(1-\lambda_{\rm min})$. The pairing state with a negative chirality 
is described by another eigenstate of the linearized GL equation as 
\begin{equation}
  \left(
  \begin{array}{c}
    \psi_{1-}({\bm r},{\bm \delta}) \\
    \psi_{2-}({\bm r},{\bm \delta}) 
  \end{array}
  \right)=
  \sum_{n\geq 0}\left(
  \begin{array}{c}
    b_{4n+2} \varphi_{4n+2}({\bm r},{\bm \delta}) \\
    b_{4n} \varphi_{4n}({\bm r},{\bm \delta})
  \end{array}
  \right). 
  \label{eq:second_eigenstate}
\end{equation} 
We here denote the n-th Landau level wave functions
\begin{equation}
\varphi_{n}({\bm r},{\bm 0})=
\sum_{m} c_{m}e^{{\rm i}mqy}(2^n\sqrt{\pi}n!)^{-1/2} H_{n}(x+mq)e^{-(x+mq)^{2}/2}
\nonumber
\end{equation}
and $\varphi_{n}({\bm r},{\bm \delta})=
e^{-{\rm i}\delta_{\rm x}(y-\delta_{\rm y})}\varphi_{n}({\bm r}-{\bm \delta},{\bm 0})$ 
using the unit of length $l=\sqrt{\varPhi_0/2\pi H}$ and the Hermite polynomials $H_{n}(x)$. 
We here assume the square lattice structure of the vortex
in accordance with the small-angle neutron scattering experiment\cite{Riseman_et_al}.
Thus, we take $c_{m}=1$ for all integers $m$ and $q=\sqrt{2\pi}$.

We adopt variational wave functions consisting of a linear combination of 
the above solutions of the linearized GL equation: 
\begin{gather}
  \left(
  \begin{array}{c}
    \Delta_{\uparrow\uparrow,1}({\bm r}) \\
    \Delta_{\uparrow\uparrow,2}({\bm r})
  \end{array}
  \right)
  =C_{1}
  \left(
  \begin{array}{c}
    \psi_{1+}({\bm r},{\bm \delta_1}) \\
    \psi_{2+}({\bm r},{\bm \delta_1}) 
  \end{array}
  \right)
  +C_{2}
  \left(
  \begin{array}{c}
    \psi_{1-}({\bm r},{\bm \delta_2}) \\
    \psi_{2-}({\bm r},{\bm \delta_2}) 
  \end{array}
  \right), 
  \label{eq:trial_fu_1}\\
  \left(
  \begin{array}{c}
    \Delta_{\downarrow\downarrow,1}({\bm r})\\
    \Delta_{\downarrow\downarrow,2}({\bm r})
  \end{array}
  \right)
  =C_{3}
  \left(
  \begin{array}{c}
    \psi_{1+}({\bm r},{\bm \delta_3}) \\
    \psi_{2+}({\bm r},{\bm \delta_3}) 
  \end{array}
  \right)
  +C_{4}
  \left(
  \begin{array}{c}
    \psi_{1-}({\bm r},{\bm \delta_4}) \\
    \psi_{2-}({\bm r},{\bm \delta_4}) 
  \end{array}
  \right), 
  \label{eq:trial_fu_2}
\end{gather}
where $C_1$ and $C_2$ ($C_3$ and $C_4$) represent Cooper pairs with up spin (down spin) 
and dominantly positive and negative chiralities, respectively. 
Two-dimensional vectors ${\bm \delta_j}$ determine the position of vortex cores. 
We take ${\bm \delta_1} = {\bm 0}$ without any loss of generality. 
These variational wave functions precisely reproduce the pairing state 
near the transition temperature, and their validity at low temperatures 
has been justified by the calculation for the chiral superconducting state\cite{Kita_PRL83_1846}. 
Substituting eqs. (\ref{eq:smallest_eigenstate})-(\ref{eq:trial_fu_2})
into eqs.~(\ref{eq:f0})-(\ref{eq:fso2}) and
integrating over the unit cell,
we obtain the free-energy density as
\begin{equation}
  F(C_1,C_2,C_3,C_4,{\bm \delta_2},{\bm \delta_3},{\bm \delta_4})=
  \frac{1}{S_{\rm c}}\int\!\! f({\bm r})\!~{\rm d}^2r,
\end{equation}
where $S_{\rm c}$ is the area of a unit cell. 
Variational parameters $(C_1,C_2,C_3,C_4,{\bm \delta_2},{\bm \delta_3},{\bm \delta_4})$ are
optimized to minimize the GL free-energy density.
We find that ${\bm \delta_4} = {\bm 0}$ and ${\bm \delta_2} = {\bm \delta_3}$ 
in the following results. 

In the following, we choose the parameters $\beta_1=1.0,~\beta_2=\beta_3=0.5,~
\xi_1/\xi=1.2,~\xi_2/\xi=0.83$, and $\xi_3/\xi=0.3$ using $\xi\equiv\sqrt{\xi_1\xi_2}$. 
For spin-orbit couplings, we assume $\epsilon=-1.5\times10^{-2}$ and $\delta=3.75\times10^{-3}$ 
unless explicitly mentioned otherwise. 
We assume the sign $\epsilon<0$ and $\delta>0$ so that a helical state with the d-vector 
${\bm d} = p_{\rm x}{\hat x}-p_{\rm y}{\hat y}$ is stabilized at zero magnetic field. 
Later, we will discuss the other cases and show that the following results are not altered
by the sign of $\epsilon$ and $\delta$. 

\begin{figure}[htbp]
  \centering
  \includegraphics[width=5.5cm,height=5.5cm,origin=c,keepaspectratio,clip]{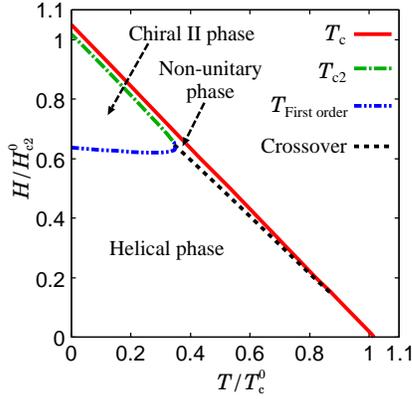}
  \hspace{0.0cm} 
  \caption{(Color online) The $H$-$T$ phase diagram of four-component GL model [eq.~(1)]
    for $\beta_1=1.0,~\beta_2=\beta_3=0.5,~\xi_1/\xi=1.2,~\xi_2/\xi=0.83,~\xi_3/\xi=0.3,~
    \epsilon=-1.5\times10^{-2}$, and $\delta=3.75\times10^{-3}$. 
    The unit of magnetic field is $H^{0}_{\rm c2}=\varPhi_0/2\pi\xi^2$. 
    The red solid line shows the superconducting transition temperature $T_{\rm c}$, 
    while the green dot-dashed line shows the second superconducting transition temperature $T_{\rm c2}$. 
    The blue double-dot-dashed line and black dashed line depict the first-order transition 
    and crossover, respectively. We define the crossover temperatures as $|C_1|=2|C_4|$.
    The non-unitary phase, helical phase, and chiral II phase depicted in the phase diagram 
    are approximately described by the d-vector 
    ${\bm d}\sim({\hat x}+{\rm i}{\hat y})(p_{\rm x}+{\rm i}p_{\rm y})$, 
    ${\bm d}\sim p_{\rm x}{\hat x}-p_{\rm y}{\hat y}$, 
    and ${\bm d}\sim(p_{\rm x}+{\rm i}p_{\rm y}){\hat x}$, respectively, 
    for our choice of parameters $\xi_1>\xi_2,~\epsilon<0$, and $\delta>0$. 
    The other cases are summarized in Tables I and II.} 
  \label{fig:diagram01} 
\end{figure}

Figure~\ref{fig:diagram01} shows the result of the $H$-$T$ phase diagram.
We see that the helical state is stable at low magnetic fields, while 
the ``chiral II state'' is stabilized for $H/H^{0}_{\rm c2}>0.62$. 
The d-vector of the chiral II state is approximately described as 
${\bm d} \sim (p_{\rm x}+{\rm i}p_{\rm y}){\hat x}$, or ${\bm d}\sim(p_{\rm x}+{\rm i}p_{\rm y}){\hat y}$, 
which is different from ${\bm d}=(p_{\rm x}\pm{\rm i}p_{\rm y})\hat{z}$ in the chiral state. 
It is shown that the superconducting double transitions occur at high magnetic fields,
the chiral II state changes to the non-unitary state near the transition temperature. 
This non-unitary state changes to the helical state at low magnetic fields through the crossover. 
The chiral state ${\bm d}=(p_{\rm x}\pm{\rm i}p_{\rm y})\hat{z}$ may be stabilized near the 
zero magnetic field\cite{Mackenize_et_al,Maeno_et_al_jpsj}, 
but we do not show this in Fig.~\ref{fig:diagram01}. 
\begin{figure}[htbp] 
  \centering
  \hspace{0.0cm}
  \includegraphics[width=8.5cm,height=4.5cm,origin=c,keepaspectratio,clip]{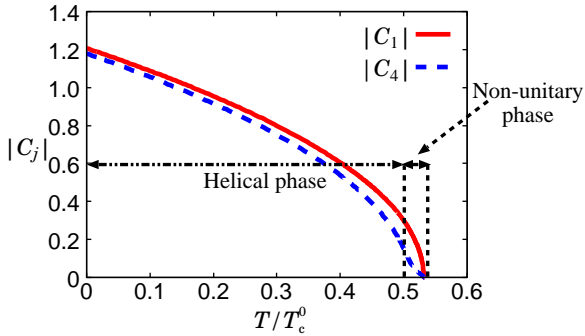}
  \caption{(Color online) Temperature dependence of the absolute values of  
    variational parameters at $H/H^{0}_{\rm c2}=0.5$. 
    The other parameters are the same as those in Fig.~1. 
    We show $|C_1|$ (red solid line) and $|C_4|$ (blue dashed line), since $C_2 = C_3 =0$.
    We define the crossover temperature as $|C_1|=2|C_4|$.}
  \label{fig:H=0.5}
\end{figure}
\begin{figure}[htbp]
  \centering
  \hspace{0.0cm}
  \includegraphics[width=8.5cm,height=4.5cm,origin=c,keepaspectratio,clip]{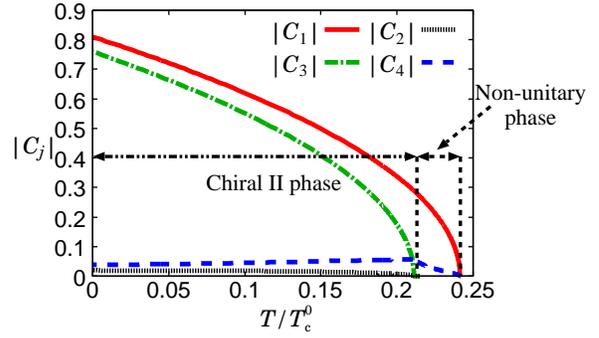}
  \caption{(Color online)
    Temperature dependence of the absolute values of 
    variational parameters $|C_1|$ (red solid line), $|C_2|$ (black dotted line), 
    $|C_3|$ (green dot-dashed line), and 
    $|C_4|$ (blue dashed line) at $H/H^{0}_{\rm c2}=0.8$. 
    The other parameters are the same as those in Fig.~1.}
  \label{fig:H=0.8}
\end{figure}
\begin{figure}[htbp]
  \centering
  \includegraphics[width=8.5cm,height=6.0cm,origin=c,keepaspectratio,clip]{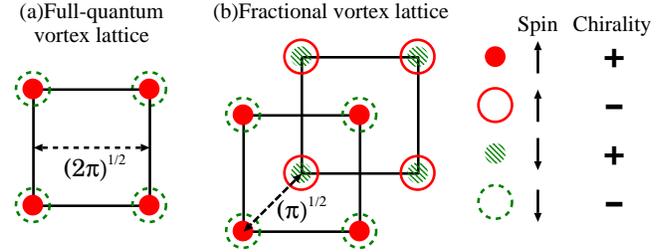}
  \hspace{0.0cm}
  \caption{(Color online) Schematic figure of vortex lattices. 
    (a)~Full-quantum vortex lattice in the non-unitary and helical phases. 
    (b)~Fractional vortex lattice in the chiral II phase. 
    Vortex cores of $C_1$ (red closed circles), $C_2$ (red open circles), 
    $C_3$ (green shaded circles), 
    and $C_4$ (green open circles with dashed lines) terms are depicted. 
    The spin and chirality of each component are shown on the right-hand side of figures. 
    We choose the unit of length $l=\xi\sqrt{H^{0}_{\rm c2}/H}$.}
  \label{fig:vortex_lattice}
\end{figure}

To clarify these pairing states, we show the temperature dependence of 
variational parameters. Figure~\ref{fig:H=0.5} shows the absolute values $|C_{j}|$ 
at an intermediate magnetic field $H/H^{0}_{\rm c2}=0.5$.  
We see that $C_1$ and $C_4$ are finite, but $C_2= C_3=0$ in the helical state 
and non-unitary state. 
Although $C_1$ and $C_4$ have nearly the same magnitude in the helical state, 
$|C_{4}| \ll |C_1|$ near the transition temperatures, indicating  
the non-unitary spin-triplet superconducting state with 
${\bm d}\sim({\hat x}+{\rm i}{\hat y})(p_{\rm x}+{\rm i}p_{\rm y})$.  
Later, we will show that the other non-unitary state
${\bm d}\sim({\hat x}-{\rm i}{\hat y})(p_{\rm x}+{\rm i}p_{\rm y})$ 
is stabilized for a positive spin-orbit coupling $\epsilon$. 
Note that the non-unitary state is stabilized 
by the spin-orbit coupling and orbital pair breaking effect in this case.
This mechanism of stabilizing the non-unitary state is different from the conventional one,
which realizes the $A_1$ phase of superfluid $^{3}$He~\cite{Leggett} 
and possibly Sr$_2$RuO$_4$.\cite{Udagawa_Yanase_Ogata} 
We elucidate this "orbital-induced non-unitary state" by looking at the chirality in 
each spin component of Cooper pairs. 
For ${\bm d} = p_{\rm x}{\hat x}-p_{\rm y}{\hat y}$, the Cooper pairs formed by quasiparticles 
with up spin have the positive chirality as $\Delta_{\uparrow\uparrow} = p_{\rm x}+{\rm i}p_{\rm y}$,
while the chirality is negative for down spin as $\Delta_{\downarrow\downarrow} = -(p_{\rm x}-{\rm i}p_{\rm y})$.
When the magnetic field is switched on, the positive chirality is enhanced by the coupling of 
chirality and magnetic field.\cite{Agterberg}
Thus, the dominant order parameter is $\Delta_{\uparrow\uparrow} = p_{\rm x}+{\rm i}p_{\rm y}$ 
in the non-unitary state, although a subdominant order parameter
$\Delta_{\downarrow\downarrow} = -(p_{\rm x}-{\rm i}p_{\rm y})$ is induced by the spin-orbit coupling.
Note that the cooperation of the chirality, magnetic field, and spin-orbit coupling plays 
an essential role in stabilizing this orbital-induced non-unitary state.

Figure~\ref{fig:H=0.8} shows the variational parameters $|C_{j}|$ 
at a rather high magnetic field $H/H^{0}_{\rm c2}=0.8$. 
We see that the parameter $C_3$ has a large magnitude in the low-temperature chiral II phase. 
This is because the magnetic field favors the positive chirality of Cooper pairs 
with down spin $\Delta_{\downarrow\downarrow} = p_{\rm x}+{\rm i}p_{\rm y}$. 
This coupling of chirality and magnetic field competes with the spin-orbit coupling,
which stabilizes the negative chirality of $\Delta_{\downarrow\downarrow}$. 
The former is more important than the latter at high fields, and therefore stabilizes the 
chiral II phase. 
On the other hand, the negative chirality component 
$\Delta_{\downarrow\downarrow} = -(p_{\rm x}-{\rm i}p_{\rm y})$, which is represented by $C_4$, 
is suppressed in the chiral II state.
Taking the $C_1$ and $C_3$ terms into account, the chiral II state is 
described by the d-vector ${\bm d}\sim(p_{\rm x}+{\rm i}p_{\rm y}){\hat x}$ or 
${\bm d}\sim(p_{\rm x}+{\rm i}p_{\rm y}){\hat y}$.
These two pairing states are degenerate because the spin-orbit coupling terms
between $C_1$ and $C_3$ cancel out. 
In other words, Cooper pairs with up spin $\Delta_{\uparrow\uparrow} = p_{\rm x}+{\rm i}p_{\rm y}$ 
are almost decoupled from those with down spin $\Delta_{\downarrow\downarrow} = p_{\rm x}+{\rm i}p_{\rm y}$ 
in the chiral II phase.
This unusual decoupling leads to the fractional vortex lattice, which we discuss below. 
\begin{table*}[t]
  \begin{center}
    \begin{tabular}{ccccc}
      \hline
      &\multicolumn{2}{c}{$\xi_1>\xi_2$}  &  \multicolumn{2}{c}{$\xi_1<\xi_2$} \\ \hline
      Chiral II phase   &\multicolumn{2}{c}{$(p_{\rm x}+{\rm i}p_{\rm y}){\hat x},~(p_{\rm x}+{\rm i}p_{\rm y}){\hat y}$}
      &\multicolumn{2}{c}{$(p_{\rm x}-{\rm i}p_{\rm y}){\hat x},~(p_{\rm x}-{\rm i}p_{\rm y}){\hat y}$}\\
      \hline
      &  $\epsilon>0$       & $\epsilon<0$    &   $\epsilon>0$   & $\epsilon<0$ \\ \hline 
      Non-unitary phase &$({\hat x}-{\rm i}{\hat y})(p_{\rm x}+{\rm i}p_{\rm y})$&
      $({\hat x}+{\rm i}{\hat y})(p_{\rm x}+{\rm i}p_{\rm y})$&
      $({\hat x}+{\rm i}{\hat y})(p_{\rm x}-{\rm i}p_{\rm y})$&
      $({\hat x}-{\rm i}{\hat y})(p_{\rm x}-{\rm i}p_{\rm y})$ \\ \hline
    \end{tabular}
  \end{center}
\caption{Summary of d-vector in the non-unitary phase and chiral II phase. 
  The magnitude relation between $\xi_1$ and $\xi_2$ determines the chirality. 
  The spin component in the non-unitary phase is determined by the sign of spin-orbit coupling $\epsilon$.} 
\label{ch_nonuni}
\end{table*}
\begin{table*}[!t]
  \begin{center}
    \begin{tabular}{ccccc}
      \hline
      &\multicolumn{2}{c}{$\epsilon>0$}&\multicolumn{2}{c}{$\epsilon<0$}\\ \hline
      &$\delta>0$       & $\delta<0$   &   $\delta>0$      & $\delta<0$ \\ \hline
      Helical phase  &$p_{\rm x}{\hat x}+p_{\rm y}{\hat y}$ & $p_{\rm y}{\hat x}-p_{\rm x}{\hat y}$
      &$p_{\rm x}{\hat x}-p_{\rm y}{\hat y}$ & $p_{\rm y}{\hat x}+p_{\rm x}{\hat y}$ \\ \hline
    \end{tabular}
  \end{center}
  \caption{Summary of d-vector in the helical phase. 
    The sign of spin-orbit couplings $\epsilon$ and $\delta$ determines the d-vector.}
  \label{heli}
\end{table*}

We now turn to the vortex lattice structure. 
Interestingly, we find that the fractional vortex lattice 
appears in the chiral II phase in contrast to the conventional full-quantum vortex lattices 
in the non-unitary and helical phases. 
We schematically illustrate these vortex lattices in Fig.~\ref{fig:vortex_lattice}. 
It is shown that the vortex cores of order parameters for 
$\Delta_{\uparrow\uparrow} = p_{\rm x}-{\rm i}p_{\rm y}$ ($C_2$) and for 
$\Delta_{\downarrow\downarrow} = p_{\rm x}+{\rm i}p_{\rm y}$ ($C_3$) 
shift from those of $\Delta_{\uparrow\uparrow}=p_{\rm x}+{\rm i}p_{\rm y}$ ($C_1$) 
and $\Delta_{\downarrow\downarrow}=p_{\rm x}-{\rm i}p_{\rm y}$ ($C_4$). 
Thus, a core-less vortex is stabilized in the chiral II phase, which is regarded as 
a fractional vortex lattice state\cite{Chung}. 
Note that this fractional vortex lattice is stabilized by the spin-orbit coupling, 
in sharp contrast to the usual role of spin-orbit coupling. 
When we assume a strong spin-orbit coupling, the fractional vortex lattice is destabilized so that 
the spin-orbit coupling energy is gained\cite{Chung}.
Contrary to this common knowledge, the role of spin-orbit coupling is altered in the chiral II phase 
because of the chirality arising from two orbital components. 
We will show details of the fractional vortex lattice structure in another paper.

Here, we investigate the parameter dependences of the phase diagram.
We would like to stress again that the structure of the phase diagram is not altered by  
the signs of the spin-orbit couplings $\epsilon$ and $\delta$ or by the magnitude
relation $\xi_1 - \xi_2$.
When we change the sign of these parameters, the d-vector in each phase changes
as summarized in Tables \ref{ch_nonuni} and \ref{heli}.
The magnitude relation between $\xi_1$ and $\xi_2$ determines the chirality (Table \ref{ch_nonuni}).
On the other hand, the signs of the spin-orbit couplings $\epsilon$ and $\delta$ determine
the d-vector of the helical phase (Table \ref{heli}).
The spin component in the non-unitary phase is also determined by the sign of $\epsilon$ 
(Table \ref{ch_nonuni}).

\begin{figure}[htbp]
  \centering
  \includegraphics[width=5.5cm,height=5.5cm,origin=c,keepaspectratio,clip]{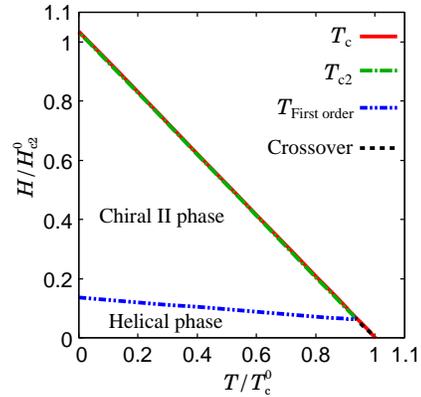}
  \hspace{0.0cm}
  \caption{(Color online) $H$-$T$ phase diagram for $\epsilon=-1.5\times 10^{-3}$ and 
    $\delta=3.75\times 10^{-4}$. 
    The other parameters are the same as those in Fig.~1.}
  \label{fig:diagram02}
\end{figure}

\begin{figure}[htbp]
  \centering
  \includegraphics[width=5.5cm,height=5.5cm,origin=c,keepaspectratio,clip]{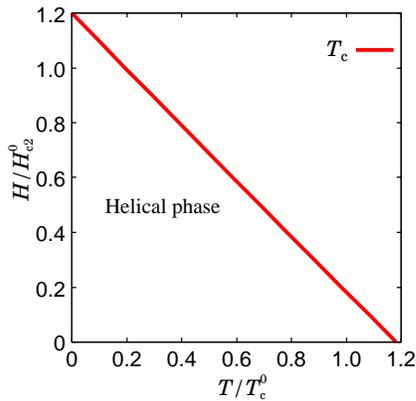}
  \hspace{0.0cm}
  \caption{(Color online) $H$-$T$ phase diagram for $\epsilon=-0.15$ and $\delta=0.0375$.
    The other parameters are the same as those in Fig.~1.}
  \label{fig:diagram03}
\end{figure}

When the magnitude of spin-orbit couplings is decreased, the chiral II phase is stabilized. 
Figure~\ref{fig:diagram02} shows that the chiral II phase is stable in almost the entire
parameter range of the $H$-$T$ phase diagram for tiny spin-orbit couplings,  
$\epsilon=-1.5\times 10^{-3}$ and $\delta=3.75\times 10^{-4}$. 
On the other hand, moderate spin-orbit couplings, $\epsilon=-0.15$ and $\delta=0.0375$, 
stabilize the helical phase in the entire parameter range, as shown in Fig.~\ref{fig:diagram03}. 
In both cases, the vortex lattice structure is not altered from Fig.~\ref{fig:vortex_lattice}.
A microscopic calculation based on the three-orbital Hubbard model 
has estimated these parameters as $\epsilon, \delta \leq 0.01$\cite{Yanase_Ogata}.  
Thus, the moderate spin-orbit couplings assumed in Fig.~\ref{fig:diagram03} are not likely realized 
in Sr$_2$RuO$_4$. 

Finally, we discuss the pairing state in Sr$_2$RuO$_4$ on the basis of our results in 
Figs.~\ref{fig:diagram01}, \ref{fig:diagram02}, and \ref{fig:diagram03}. 
Until now, double superconducting transitions have not been observed in Sr$_2$RuO$_4$ 
in the magnetic field along the {\it c}-axis\cite{Mackenize_et_al,Maeno_et_al_jpsj}. 
This experimental status implies tiny spin-orbit couplings as in Fig.~\ref{fig:diagram02} 
or moderate spin-orbit couplings as in Fig.~\ref{fig:diagram03}. 
The chiral II state is realized in the former case, 
while the helical state is realized in the latter case. 
The latter interpretation is, however, incompatible with the results of the NMR experiment,
which did not show any temperature dependence of
Knight shift through $T_{\rm c}$\cite{Ishida_et_al,Murakawa_et_al}. 
On the other hand, the case of tiny spin-orbit couplings in Fig.~\ref{fig:diagram02} seems to 
be consistent with the experiments. Although the fractional vortex lattice in 
Fig.~\ref{fig:vortex_lattice}(b) has not been observed in the small-angle neutron scattering 
experiment\cite{Riseman_et_al}, this discrepancy is not a serious issue because the fractional 
vortex lattice is affected by the strong coupling effect as well as by the Meissner effect. 
In our future study, we will investigate these effects on the vortex lattice structure. 
In contrast to these cases, the multiple-phase diagram as shown in Fig.~\ref{fig:diagram01}
should appear in Sr$_2$RuO$_4$ when spin-orbit couplings are small but not tiny. 
No clear experimental observation has been reported for the multiple-phase diagram.
However, weak evidence has been obtained by a small-angle neutron scattering experiment\cite{Riseman_et_al}.
When the fractional vortex lattice is realized in the chiral II phase, the spatial distribution 
of the magnetic field is smeared, consistent with the unclear field distribution
obtained in an experiment at high magnetic fields\cite{Riseman_et_al}. 
Further experimental study is desired to determine the superconducting state of 
Sr$_2$RuO$_4$ in the {\it c}-axis magnetic field.

In summary, we studied the spin-triplet superconducting state of Sr$_2$RuO$_4$ in the magnetic
field along the {\it c}-axis. We assumed a weak spin-orbit coupling of Cooper pairs and considered 
the d-vector parallel to the {\it ab}-plane. 
We determined multiple phases against magnetic fields and temperatures by analyzing 
the four-component GL model using the variational method. 
Interestingly, we showed three spin-triplet superconducting phases, which are orbital-induced 
non-unitary phase, chiral II phase, and helical phase. 
We clarified the synergistic and competing roles of the chirality, magnetic field,
and spin-orbit coupling for these pairing states. 
We would like to stress that these intriguing superconducting phases are obtained 
because we take into account a weak but finite spin-orbit coupling and allow the mixing of 
irreducible representations in the order parameters. 
We also found that the fractional vortex lattice is stabilized by a weak spin-orbit coupling 
in the chiral II phase.
This is in sharp contrast to the theory of non-chiral spin-triplet superconductors\cite{Chung}. 
This finding pave the way for realizing a novel topological defect with Majorana fermion\cite{Ivanov} 
in spin-triplet superconductors with spin and orbital degrees of freedom. 

The authors are grateful to K. Ishida, Y. Maeno, Y. Matsuda, S. Yonezawa, and S. Kashiwaya
for fruitful discussions. 
We especially thank M. Sigrist for stimulating discussions on the spin-triplet superconductivity. 
This work was supported by KAKENHI (Grant Nos. 24740230 and 23102709).

\end{document}